\documentclass[]{spie}  

\usepackage{amsmath,amsfonts,amssymb}
\usepackage{graphicx}
\usepackage[colorlinks=true, allcolors=blue]{hyperref}
\usepackage{comment}
\usepackage{subcaption}
\usepackage{wrapfig}

\usepackage{url}
\usepackage{hanging}
\usepackage{lipsum}
\usepackage{multicol}
\usepackage{multirow}
\usepackage{overpic}
\usepackage[inline]{enumitem}
\usepackage{setspace}
\usepackage[labelfont=bf, singlelinecheck=false, justification=justified]{caption}
\usepackage{wrapfig}
\usepackage{comment}
\usepackage{multicol}
\usepackage[normalem]{ulem}
\usepackage{parskip}  
\usepackage{multirow}
\usepackage[table,xcdraw,dvipsnames]{xcolor}
\usepackage[linewidth=0.5pt,skipabove=10pt,skipbelow=10pt,leftmargin=0pt,rightmargin=0pt,framemethod=TikZ]{mdframed}
\definecolor{myblue}{RGB}{25, 118, 210}
\mdfdefinestyle{bluebox}{backgroundcolor=myblue!10,roundcorner=8pt,linecolor=myblue}
\usepackage{pdfpages}  
\usepackage{soul,color} 

\title{Developments on LLNL’s high contrast testbed and Lick/ShaneAO}

\author[a]{Benjamin L. Gerard}
\author[a]{Dominic F. Sanchez}
\author[b]{Aditya R. Sengupta}
\author[a]{Bautista R. Fernandez}
\author[a]{Cesar Laguna}
\author[b]{Christopher Ratliff}
\author[b]{Daren Dillon}
\author[c]{Sylvain Cetre}
\author[a]{David Tucker}
\author[a]{Mike Kim}
\author[a]{Lisa Poyneer}
\author[a]{Brian Bauman}
\author[b]{Elinor Gates}
\author[b]{Maureen Savage}
\author[b]{Rebecca Jensen-Clem}
\author[a]{S. Mark Ammons}
\author[b]{Phil Hinz}
\author[b]{Bruce Macintosh}

\affil[a]{Lawrence Livermore National Laboratory}
\affil[b]{Univeristy of California Santa Cruz}
\affil[c]{Wakea Consulting}

\authorinfo{Further author information. Send correspondence to Benjamin L. Gerard: \texttt{gerard3@llnl.gov}
}
\begin{document} 
\maketitle

\begin{abstract}
LLNL has recently setup a High Contrast Testbed (HCT) for AO and exoplanet imaging technology development. We present the various HCT technologies currently under development, including (1) a Wynne corrector, (2) multi-wavefront sensor (WFS) single conjugate AO (SCAO) control. We present HCT testing results of a first Wynne corrector prototype with a self-coherent camera. We present updates on development efforts to design and apply multi-WFS SCAO control to our HCT setup. We also present ongoing HCT deformable mirror and WFS upgrades. Lastly, we present developments for REDWOODS, a project to deploy many of these technologies on-sky on a sub-bench of the Shane AO system at Lick Observatory.
\end{abstract}


\section{INTRODUCTION}
\label{sec:intro}  
Advancing exoplanet imaging technologies at both current facilities and future extremely large telescopes (ELTs) has been identified as a top priority in the coming decades.\cite{NASEM2021} Such advancements are needed to see exoplanets that are below the current $\sim$2 Jupiter mass, $\sim$10 au, 100 Myr threshold\cite{Nielsen2019} with current 10m-class facilities and in principle enable detecting habitable zone exoplnaets around low mass stars with ELTs.\cite{Jensen-Clem2022}. In this proceeding we present such exoplanet imaging technology developments on a variety of ongoing activities led by LLNL, including wavefront sensing (WFS) and control in the lab (\S\ref{sec:redwoods}) and at the ShaneAO system of Lick Observatory (\S\ref{sec:redwoods}).
\section{LLNL's High Contrast Testbed}
\label{sec:hct}
\begin{figure}[!h]
    \centering
    \includegraphics[width=1.0\linewidth]{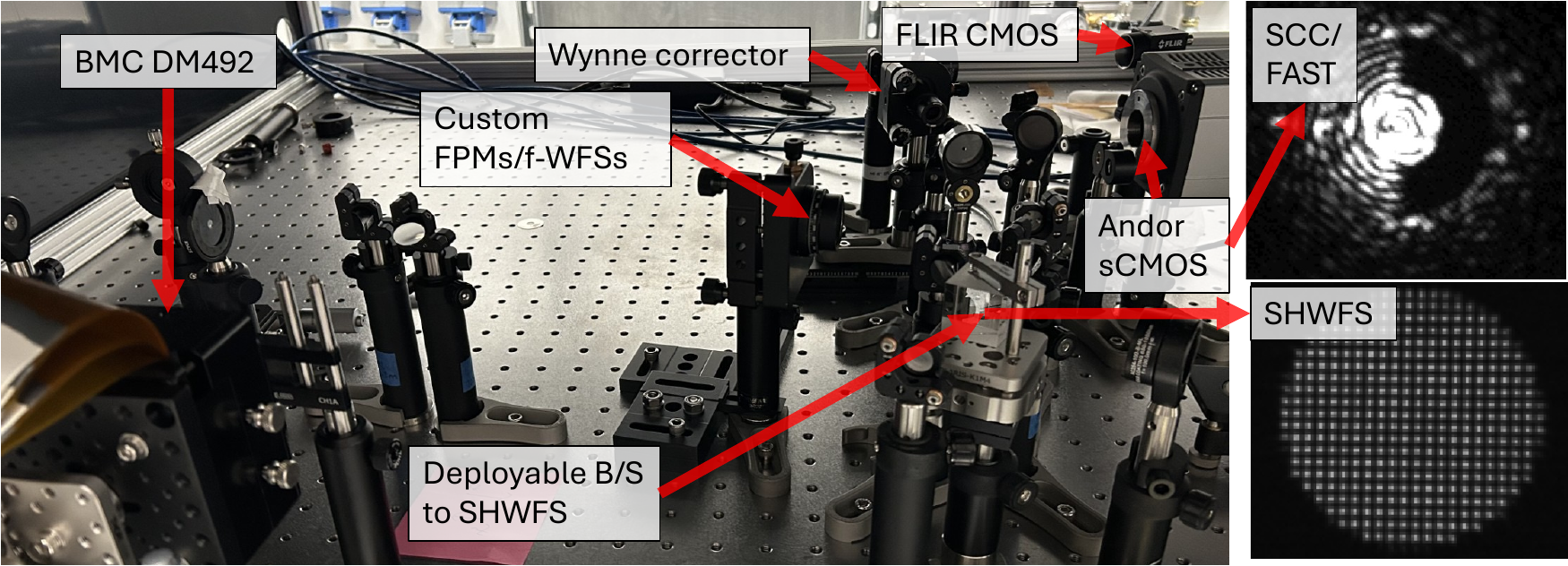}
    \caption{Illustration of LLNL's high contrast testbed (HCT), which includes a 494 actuator DM, various wavefront sensors, and $>$ 100 Hz loop speed real-time control in Python. Ongoing HCT upgrades will implement two new DMs, two new lower and higher stoke modes to the current 492 DM, more wavefront sensor options for both ground- and space-based applications.}
    \label{fig:hct}
\end{figure}

LLNL's high contrast testbed (HCT), illustrated in Fig. \ref{fig:hct}, is a on-air testbed facility developed over the past few years for the purpose of high contrast imaging (HCI) and adaptive optics (AO) technology development.
For HCT data presented in this proceeding, the testbed includes:
\begin{itemize}
\item a Boston Micromachines Corporation (BMC) 492 actuator DM with 16 bit electronics and 100V maximum voltage, providing about 700 nm of maximum inter-actuator stroke and 10 pm of minimum stroke resolution. This tradeoff of stroke reduction to increase stroke resolution was made to enable enough stroke to flatten the DM and use for more ground-relevant second-stage AO developments as well as more space-relevant pm-level wavefront control experiments.
\item a f/40 reflective focal plane with various reflective focal plane mask (FPM) and fourier-based wavefront sensor (f-WFS) mask options, with the main use at the moment being a Fast Atmospheric Self-coherent camera (SCC) Technique (FAST) FPM\cite{Gerard2018}.
\item three cameras, including a Andor Zyla 5.5 sCMOS and two Teledyne FLIR Blackfly cameras. The Andor camera is used for FAST/f-WFS via a deployable pupil imaging lens, one FLIR camera is used for a Wynne corrector developments (see \S\ref{sec:wynne}), and another is used for a quad-cell (1.1 pixels/resel) Shack Hartmann WFS (SHWFS). The SHWFS path is illuminated by a deployable beam cube.
\item Fully reflective optics from the light source to the FAST cameras apart from the DM window (wedged and AR coated for 400 - 1100 nm) to support visible spectral bandwidth tests. All beam sizes are below 7mm, enabled by the BMC DM, which allow commercial Off The Shelf (COTS) off-axis parabolic mirrors (OAPs) to produce around 39 nm rms integrated mid-to-high-spatial frequency wavefront error from OAP+fold mirror figure error. 
\item A real-time control (RTC) software framework on a 1U server, based on the UC Santa Cruz SEAL krtc framework\cite{Jensen-Clem2021}, that uses low-level shared-memory semaphores to enable high-level RTC Python scripting at $>$ 100 Hz loop speeds. The server OS is Linux RHEL 8.
\end{itemize}

Currently HCT is undergoing various upgrades, including
\begin{itemize}
\item several BMC DM upgrades (all with wedged windows with AR coatings spec'd for 400-1000 nm), including
\begin{itemize}
\item two new modes to our BMC 492 actuator DM (hereafter referred to as high order DM, or HODM): (1) a 30 nm inter-actuator stroke mode with 16 bit mode electronics to enable 0.5 pm stroke resolution for space-based AO applications, and (2) a standard 1.5 $\mu$m inter-actuator stroke with 14 bit electronics for ground-based applications. These two modes are enabled by two new electronics cards installed in the same chasis as the original HODM. A custom 3:1 multiplexer (MUX) is also being installed in a fourth slot of the same chasis to enable switching between any of the three modes completely in software.
\item a COTS BMC 140 actuator 5.5 $\mu$m stroke continuous DM with a USB (standard) driver interface (hereafter referred to as the LODM)
\item a COTS BMC 111 actuator 3.5 $\mu$m stroke hex DM with a cameralink (X) driver interface, hereafter referred to as the hex DM)
\end{itemize}
\item several new reflective scalar WFS masks designed for $\lambda_0=635$nm made by Zeiss by reactive ion etching\cite{Cumme2015} with $<$ 3\% etching depth errors, including
\begin{itemize}
    \item a tip/tilt+Gaussian FPM for FAST
    \item a Zernike WFS with r=1 $\lambda_0/D$ following Ref. \citenum{Chambouleyron2021}
    \item a charge 2 vortex FPM
    \item an anti-aliased 3 sided PyWFS, where the anti-aliasing effect is a super-Nyquist ($>$12 c/p) spatial frequency pupil in the center of the three conventional pupils used for wavefront sensing and control. 
\end{itemize}
\item an NKT super K white light laser, converted via a custom module from class 3B (COTS) to 3r for operational LLNL laser safety reasons. 
\end{itemize}

\subsection{Wavefront Control}
\label{sec:wfc}
\begin{figure}[!h]
    \centering
    \begin{subfigure}[t]{0.42\textwidth}
    \includegraphics[width=1.0\linewidth]{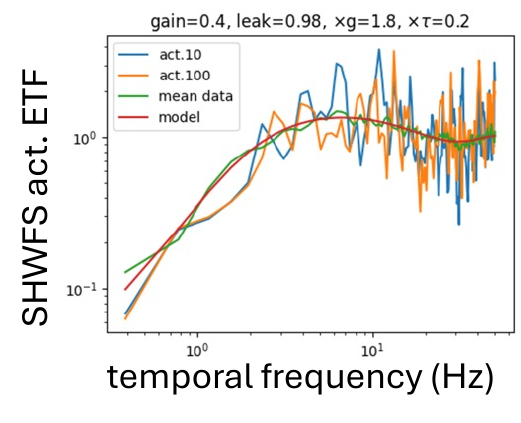}
    \caption{SHWFS ETFs in a zonal/actuator basis.}
    \end{subfigure}
    \begin{subfigure}[t]{0.57\textwidth}
    \includegraphics[width=1.0\linewidth]{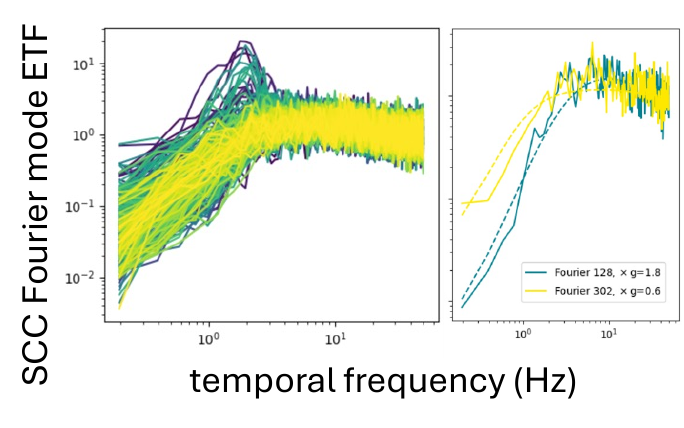}
    \caption{FAST/SCC high order Fourier mode ETFs. $\times\tau$=1.8.}
    \end{subfigure}    
    \caption{SHWFS and SCC measured error transfer functions (ETFs) and corresponding model fits running at 100 Hz frame rate. For each case the AO loop is closed on the HODM only using the given single WFS.$\times\tau$ and $\times g$ refers to the fractional delay (relative to 1 frame = 10 ms) and optical gain (separate from the control gain) produced by model ETF fits.}
    \label{fig:etfs}
\end{figure}

HCT wavefront control efforts so far have considered closed loop performance with the SHWFS and FAST WFSs at 100 Hz loop speeds, for which corresponding error transfer functions (ETFs) are shown in Fig. \ref{fig:etfs}.
SHWFS ETFs in Fig. \ref{fig:etfs}a, generated by closing the loop on-air, behave as expected, with individual actuator ETFs averaging (separately averaging open- and closed-loop PSDs) to reduce the noise for the average actuator ETF in green in good agreement with the model in red. Our SHWFS reconstructor is a matrix vactor multiply (MVM) from pixels to zonal actuator coefficients and then another MVM from coefficients to voltages. SCC/FAST ETFs in Fig. \ref{fig:etfs}b, building off the work of Ref. \citenum{Gerard2022} show a different story, illustrating a a large variation of optical gains vs. Fourier mode. Note that Low order Zernike ETFs are also measured and well-modeled but not shown in Fig. \ref{fig:etfs}. Some Fourier modes as shown show unexpected overshoot behavior and thus still require further calibration. These Fourier mode ETFs are taken on an algorithmically aligned FPM after DM raster scanning as in Ref. \cite{Gerard2022}, but not on a dark hole reference image as we found that read noise ($\sim$1e-/frame for our Andor Zyla camera) above the photon noise caused the ETFs to flatten at low temporal frequencies. Further work is still needed to sufficiently calibrate optical gains for all Fourier modes such that these modal gain variations can be "divided out" in the command matrix and produce uniform ETFs across all modes like for the SHWFS ETFs in Fig. \ref{fig:etfs}a. There is an added complication of non-linearities changing the optical gains with input disturbance amplitude, but this is beyond the scope of this proceeding.

\begin{figure}[!h]
    \centering
    \begin{subfigure}[t]{0.43\textwidth}
    \includegraphics[width=1.0\linewidth]{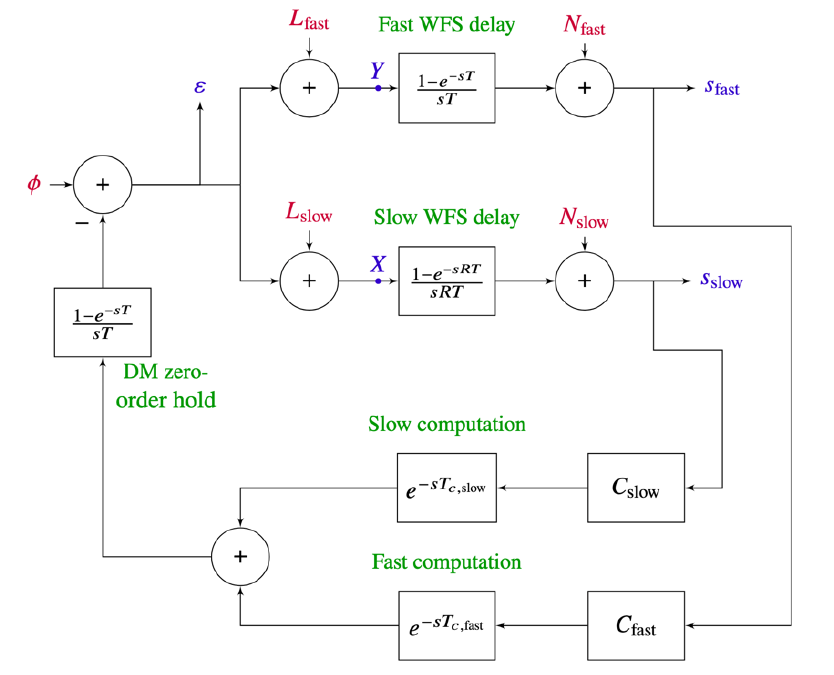}
    \caption{Control diagram, including input common path disturbances, $\phi$, temporal non-common path errors---$L_\text{fast}$ and $L_\text{slow}$---and noise---$N_\text{fast}$ and $N_\text{slow}$---on the fast and slow WFS arm, respectively.}
    \end{subfigure}
    \hfill
    \begin{subfigure}[t]{0.55\textwidth}
    \includegraphics[width=1.0\linewidth]{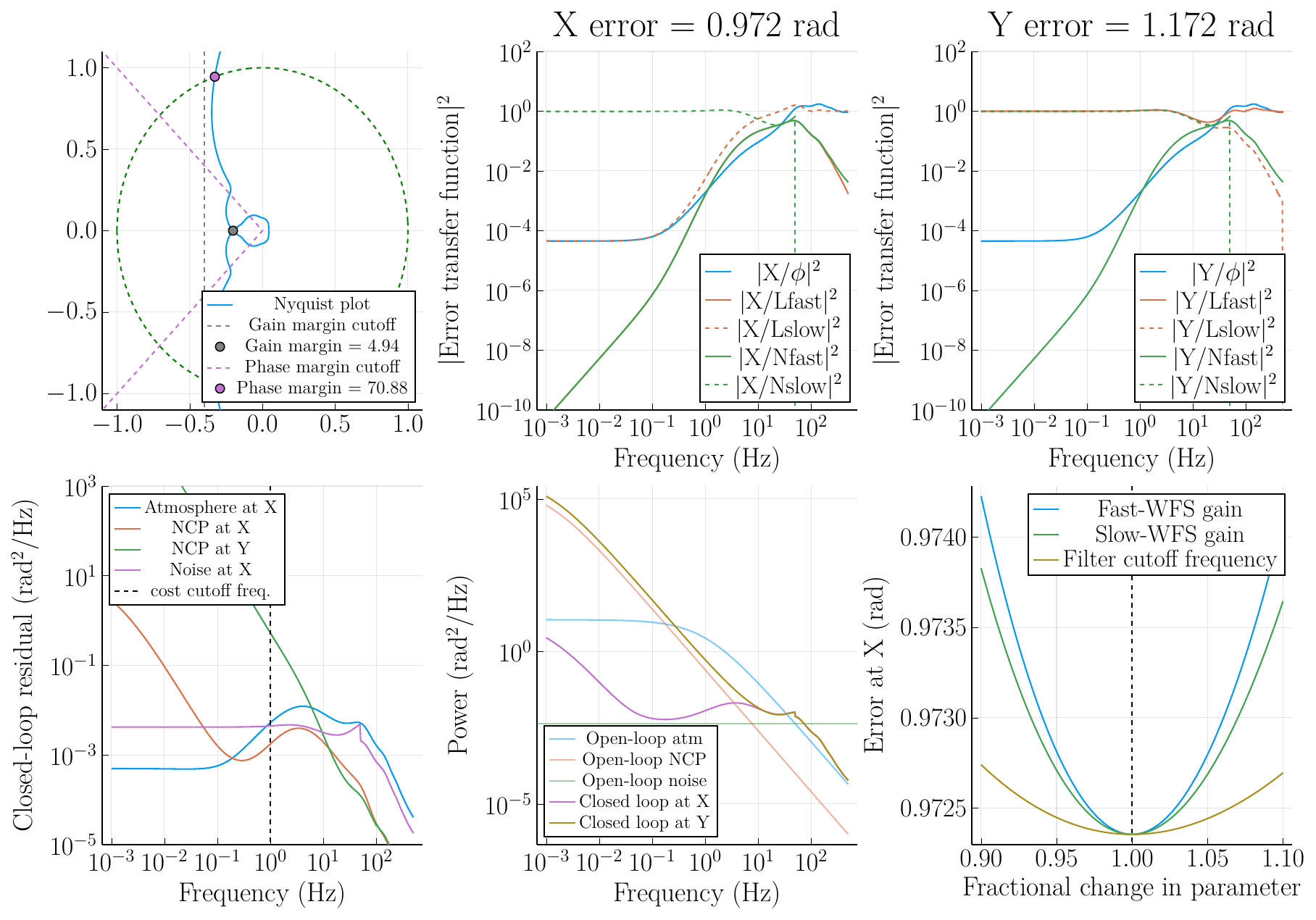}
    \caption{Results of our multi-WFS SCAO control analysis showing that an added high-pass filter to the integral controller on the fast WFS reduces inter-arm temporal NCP transfer, which we found to be needed when NCP strength is $\gtrsim$30 nm rms.}
    \end{subfigure}    
    \caption{Multi-WFS SCAO control framework and results, with more details to be presented in a future paper (Sengupta et al., in prep).}
    \label{fig:multiwfs}
\end{figure}

In tandem with HCT wavefront control developments, we have also been developing the theory for multi-WFS single conjugate AO (SCAO) control, building on Ref. \citenum{Gerard2023}. In short, a multi-WFS SCAO control framework is needed for many second-stage/focal plane WFSs whose capture range is generally diffraction-limited, thus requiring a first stage WFS, but for this scenario we consider the use of one common path DM for both WFSs. A separate paper on this topic is in preparation (Sengupta, Poyneer, Gerard et al., in prep) but brefiely summarized here in Fig. \ref{fig:multiwfs}.
In short, we found that an added high-pass filter to the integral controller (IC) on the fast WFS (keeping a conventional IC on the slow WFS arm)  reduces inter-arm temporal NCP transfer, which we found to be needed when NCP strength is $\gtrsim$30 nm rms to meet the X error $<$ 1 rad rms dynamic range requirement.
\subsection{Wynne Corrector}
\label{sec:wynne}
\begin{figure}[!h]
    \centering
    \begin{subfigure}[t]{0.5\textwidth}
    \includegraphics[width=1.0\linewidth]{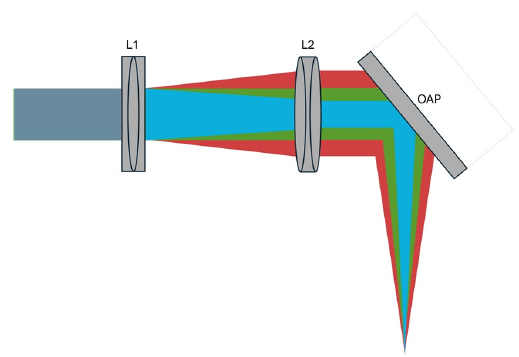}
    \caption{Our Wynne corrector design forms a linearly varying beam size with wavelength on an OAP.}
    \end{subfigure}
    \begin{subfigure}[t]{0.49\textwidth}
    \includegraphics[width=1.0\linewidth]{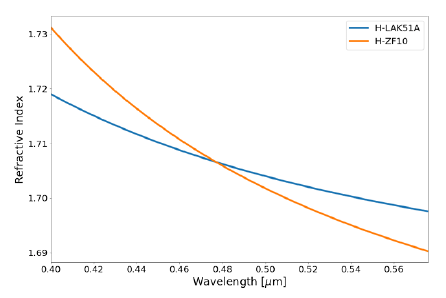}
    \caption{The center wavelength of $\lambda_0$=480nm is based on the refractive index curve crossing point of the two materials used for the two triplet lenses.}
    \end{subfigure}    
    \caption{Our Wynne corrector concept, from Ref. \citenum{Sanchez2024}, enabling a $\Delta\lambda/\lambda_0 \sim30\%$ bandwidth SCC for high-speed wavefront sensing and control of residual atmospheric speckles.}
    \label{fig:wynne_design}
\end{figure}

\begin{figure}[!h]
    \centering
    \includegraphics[width=1.0\linewidth]{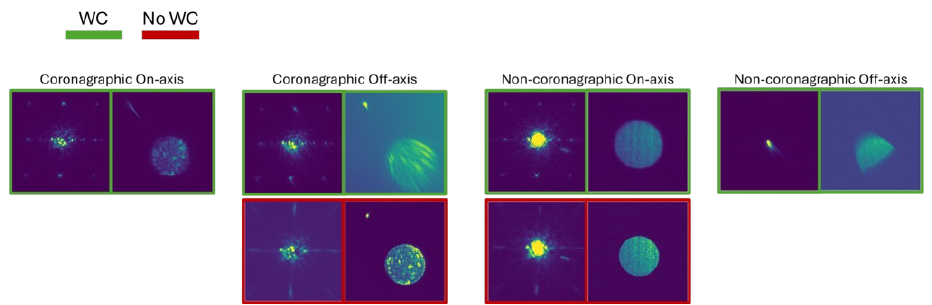}
    \caption{HCT Wynne corrector testing results. More details to come in Sanchez et al., in prep.}
    \label{fig:wynne_results}
\end{figure}

We have also developed and tested a Wynne corrector for broadband achromatic SCC operations, a concept initially proposed by Ref. \citenum{Wynne1979} and of interest to the SCC for broadband operations\cite{Galicher2007} but never tested until now. Ref. \citenum{Sanchez2024} published our design, summarized in Fig. \ref{fig:wynne_design}.
A full paper with our design and testing results is being prepared now (Sanchez, Gerard et al., in prep), but in this proceeding we show some testing results in Fig. \ref{fig:wynne_results}.
Fig. \ref{fig:wynne_results} shows measured pupil and focal images with (1) no wynne corrector (a) with and (b) without a coronagraph, shown in red, (2) an on-axis Wynne corrector (a) with and (b) without a coronagraph, and (3) an off-axis Wynne corrector (a) with and (b) without a coronagraph. HODM print through in focal plane images are clearly visible and clearly aligned when the wynne corrector is in and smeared out when the Wynne corrector is out as expected. However, the coronagraphic pupil image with the on-axis Wynne corrector illustrates a problem we did not anticipate in the design: the SCC Lyot stop pinhole is off-axis (3.4mm off-axis for our HCT setup), and the Wynne corrector exhibits significant chromatic dispersion for off-axis beam displacement well below the separations of where the SCC pinhole is. As a result, we translated our Wynne corrector assembly off-axis, improving fringe visibility but not degrading Strehl due to the relaxation of coronagraphic pupil wavefront errors, further corroborated by the off-axis non-coronagraphic Wynne image having a much worse dispersive effect to the main pupil. More details will be presented in Sanchez et al., in prep.
\section{REDWOODS}
\label{sec:redwoods}
\begin{figure}[!h]
    \centering
    \begin{subfigure}[t]{0.33\textwidth}
    \includegraphics[width=1.0\linewidth]{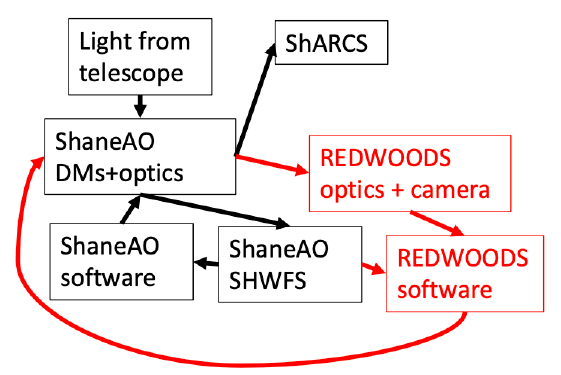}
    \caption{REDWOODS concept, including existing ShaneAO hardware and software in black and REDWOODS components in red.}
    \end{subfigure}\hfill
    \begin{subfigure}[t]{0.25\textwidth}
    \includegraphics[width=1.0\linewidth]{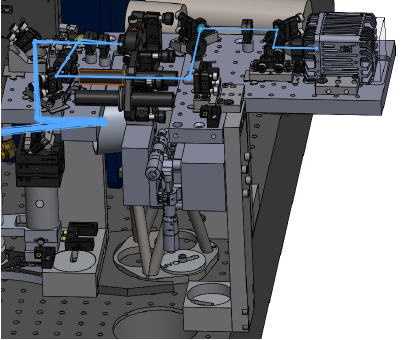}
    \caption{Opto-mechanical layout of REDWOODS (beginning with a periscope fold mirror) within ShaneAO.}
    \end{subfigure}\hfill
    \begin{subfigure}[t]{0.38\textwidth}
    \includegraphics[width=1.0\linewidth]{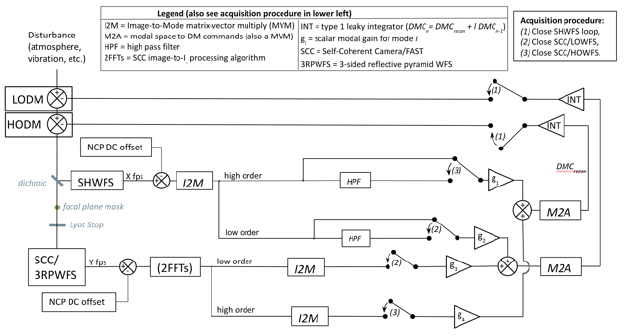}
    \caption{REDWOODS' high-level multi-WFS software design, building off of \citenum{Gerard2023}.}
    \end{subfigure}   
    \caption{REDWOODS concept (a), hardware (b) and software (c) diagrams. More details will be presented in a future paper.}
    \label{fig:redwoods}
\end{figure}

REDWOODS (Real time Exoplanet Direct imaging via Wavefront control Of Optical DefectS)---an NSF + LLNL LDRD + UCO-funded project---is an engineering demonstrator AO module of the ShaneAO system at Lick Observatory\cite{Gavel2014} that will develop and test on-sky the following new technologies, building off of LLNL HCT developments presented in \S\ref{sec:hct}:
\begin{itemize}
    \item diffraction-limited second stage wavefront sensing and control of residual atmospheric speckles at $>$100 Hz loop speeds, including
    \begin{itemize}
        \item a $lambda_0=1.05\mu$m FAST/SCC mode, including a Wynne corrector-enabled mode with $\Delta\lambda/\lambda_0\sim$20\%,
        \item a fully-reflective 3 sided near infrared pyramid WFS with $\Delta\lambda/\lambda_0\sim$50\%, building off of previous reflective 3 sided pyramid WFS developments at ShaneAO\cite{Sanchez2024b},
    \end{itemize}
    \item multi-WFS SCAO real-time control software, and
    \item kHz-speed SHWFS wavefront sensing and control using reduced intensities instead of slopes.
\end{itemize}
The REDWOODS high-level concept, opto-mechanical, and high-level real-time control diagrams are shown in Fig. \ref{fig:redwoods}.
Note that Hurtado et al. in this proceedings investigates expected performance with the REDWOODS Pyramid WFS. In April 2024 the instrument completed it pre-fabrication review and it currently in the integration and testing phase. Installation at ShaneAO/Lick is expected to be completed later this fall, with internal source testing and then on-sky testing expected thereafter through Sep. 2026. More details about REDWOODS hardware and software design and testing will be presented in a future paper later this fall. 
\section{CONCLUSION}
\label{sec:conclusion}
Exoplanet imaging developments at LLNL, in collaboration with UCSC, are pushing a new generation of AO technologies with the goal of bridging the $\gtrsim100\times$gap to enable habitable exoplanet imaging with future ground- and space-based observatories\cite{Jensen-Clem2022,Ganel2022}. Several such technology development efforts are presented in this proceedings, including LLNL's new high contrast testbed (HCT) facility (\S\ref{sec:hct}), corresponding experimental and theoretical wavefront sensing and control developments on this testbed related to multi-wavefront sensor single conjugate AO control (\S\ref{sec:wfc}) and high-spectral bandwidth focal plane wavefront sensing with a self-coherent camera (\S\ref{sec:wynne}), and a new AO demonstrator module of ShaneAO called REDWOODS being developed to test many of these technologies on-sky in the next $\sim$year (\S\ref{sec:conclusion}). 
\acknowledgments 
This work was performed under the auspices of the U.S. Department of Energy by Lawrence Livermore National Laboratory under Contract DE-AC52-07NA27344. This document number is LLNL-PROC-2010116.
%
\bibliography{report} 

\begin{thebibliography}{10}

\bibitem{NASEM2021}
NASEM,  [{\em Pathways to Discovery in Astronomy and Astrophysics for the 2020s}{\nolinebreak\hspace{0.1em}]}, The National Academies Press (2021).

\bibitem{Nielsen2019}
Nielsen+, ``The gemini planet imager exoplanet survey: Giant planet and brown dwarf demographics from 10 to 100 au,'' {\em The Astronomical Journal}~{\bf 158},  13 (2019).

\bibitem{Jensen-Clem2022}
Jensen-Clem+, ``A technology and science gap list for habitable-zone exoplanet imaging with ground-based extremely large telescopes,'' {\em Adaptive Optics Systems VIII}~{\bf 12185},  1218503 (2022).

\bibitem{Gerard2018}
Gerard+, ``Fast coherent differential imaging on ground-based telescopes using the self-coherent camera,'' {\em The Astronomical Journal}~{\bf 156},  106 (2018).

\bibitem{Jensen-Clem2021}
Jensen-Clem+, ``The santa cruz extreme ao lab (seal): design and first light,'' {\em Techniques and Instrumentation for Detection of Exoplanets X}~{\bf 11823},  118231D (2021).

\bibitem{Cumme2015}
Cumme+, ``From regular periodic micro-lens arrays to randomized continuous phase profiles,'' {\em Advanced Optical Technologies}~{\bf 4},  47 (2015).

\bibitem{Chambouleyron2021}
Chambouleyron+, ``Variation on a zernike wavefront sensor theme: Optimal use of photons,'' {\em Astronomy and Astrophysics}~{\bf 650},  L8 (2021).

\bibitem{Gerard2022}
Gerard+, ``Laboratory demonstration of real-time focal plane wavefront control of residual atmospheric speckles,'' {\em Journal of Astronomical Telescopes, Instruments, and Systems}~{\bf 8},  039001 (2022).

\bibitem{Gerard2023}
Gerard+, ``First laboratory demonstration of real-time multi-wavefront sensor single conjugate adaptive optics,'' {\em Society of Photo-Optical Instrumentation Engineers (SPIE) Conference Series}~{\bf 12680},  126801Q (2023).

\bibitem{Wynne1979}
Wynne+, ``Extending the bandwidth of speckle interferometry.,'' {\em Optics Communications}~{\bf 28},  21 (1979).

\bibitem{Galicher2007}
Galicher+, ``Expected performance of a self-coherent camera,'' {\em Comptes Rendus Physique}~{\bf 8},  333 (2007).

\bibitem{Sanchez2024}
Sanchez+, ``Developing a wynne corrector for higher spectral bandwidth focal plane wavefront sensing,'' {\em Advances in Optical and Mechanical Technologies for Telescopes and Instrumentation VI}~{\bf 13100},  131004Q (2024).

\bibitem{Gavel2014}
Gavel+, ``Shaneao: wide science spectrum adaptive optics system for the lick observatory,'' {\em Adaptive Optics Systems IV}~{\bf 9148},  914805 (2014).

\bibitem{Sanchez2024b}
Sanchez+, ``A compact three-sided reflective pyramid wavefront sensor: optical design and closed-loop adaptive optics testbed characterization,'' {\em Adaptive Optics Systems IX}~{\bf 13097},  130971R (2024).

\bibitem{Ganel2022}
Ganel+, ``Astrophysics strategic technology gaps following the 2020 decadal survey,'' (2022).
\newblock \href{https://science.nasa.gov/astrophysics/programs/astrophysics-division-technology/technology-gaps/#priorities}{link}.

\end{thebibliography}
\bibliographystyle{spiebib} 
\end{document}